\PassOptionsToPackage{numbers}{natbib}
\input{aipcheck}

\documentclass[
    ,final            
  ]
  {aipproc}

\layoutstyle{8x11single}

\begin{document}

\title{Strange Baryonic Resonances and Resonances Coupling to Strange Hadrons at SIS Energies}


\author{L. Fabbietti for the HADES Collaboration }{
  address={e12, Physik Department Technische Universit\"at M\"unchen\\
  Excellence Cluster "Origin and Structure of the Universe" }
}

%

\begin{abstract}
 The role played by baryonic resonances in the production of final states containing strangeness for proton-proton
 reactions at 3.5 GeV measured by HADES is discussed by means of several very different measurements.
 First the associate production of $\Delta$ resonances accompanying final states with strange hadrons
 is presented, then the role of interferences among N$^*$ resonances, as measured by HADES
 for the first time, is summarised.
 Last but not least the role played by heavy resonances, with a mass larger than $2$ GeV/c$^2$ in the 
 production of strange and non-strange hadrons is discussed. Experimental evidence for the presence
 of a $\Delta(2000)^{++}$ are presented and hypotheses are discussed employing the contribution
 of similar objects to populate the excesses measured by HADES for the $\Xi$ in A+A and p+A collisions
 and in the dilepton sector for A+A collisions.
 This extensive set of results helps to better understand the dynamic underlaying particle production in elementary
 reactions and sets a more solid basis for the understanding of heavy ion collisions at the same energies and even higher
 as planned at the FAIR facility.
\end{abstract}

\maketitle


\section{Introduction}
The role played by resonances in hadron production in elementary and heavy ions
collisions in the GeV energy range turned out to be fundamental even at a kinetic energy
of 3.5 GeV. Previously it has been indeed thought that around $\sqrt{s}=\,3$ GeV the contribution by baryonic
resonances should become less important. The HADES collaboration has published in the last
three years several measurements that confirm the importance of the baryonic resonances 
associated to or leading to strangeness production providing new information for some exclusive
production channels. Hence a first aspect consists in the quantitative determination
of the resonance production associated to a strange ( or non strange) final state.
As a matter of fact all the models used to interpret the available data in p+p, p+A and A+A 
collisions do not include any interferences among the possible contributing resonances
and in this work we show a clear example where the interference pattern is very well recognisable 
and hence not negligible. So not only resonances join the production of strange hadrons,
but different resonances decaying into the same final state may interfere and the cross sections
and angular distribution will be influenced by this process. This second aspect also represents
one of the fundamental questions of our field at the moment.
The dynamics of particle production in elementary collisions is clearly interesting on its own
but it also represents an important and necessary reference to interpret the experimental
data collected for heavy ion reaction, where in the GeV energy range matter is supposed to 
get highly compressed during the collisions and hence new properties of the hadrons
within this compressed matter might become visible.
For this reason, a third aspect of the resonance study is the sequential decay
 of different resonances
starting with rather massive objects and their production probability in elementary and heavy ion collisions.
 The excitation of heavy resonances with respect to the
available phase space for a given reaction has been not considered up to now in the interpretation of the
data for p+p, p+A and A+A reactions in the GeV region, and aside the strangeness sector the population
of these heavy resonances might also modify the measured yield of non-strange observables.
The question which can be asked in this context is if at these energies the observed excesses for dilepton in the intermediate
mass range or the production of the $\Xi$ in heavy ion collisions might also be linked to the decay
of massive resonances.
In the following these three different aspects of the resonance production  results will be discussed on the base
of the recent HADES.
 
\section{Associate Production of Resonances to Strange Hadrons}
We begin with considering the investigation of the $\Lambda(1405)$ resonance produced in 
p+p collisions at $3.5$ GeV \cite{L1,L2}. Here the reaction $p+p\rightarrow \Lambda(1405)(\rightarrow 
\Sigma^{\pm}+\pi^{\mp})+p+K^+$  
is analysed considering also the successive decay $\Sigma^{\pm} \rightarrow \pi^{\pm} +n$.
\begin{figure}
  \includegraphics[height=.30\textheight]{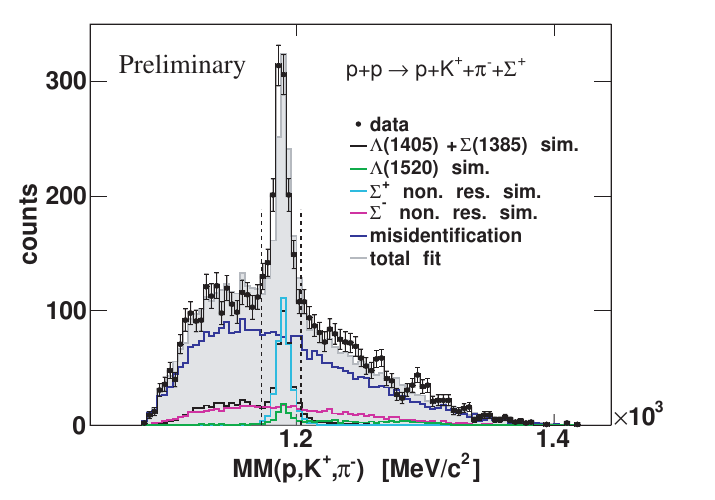}
  \caption{(Color online) Missing mass to ($p,\,K^+,\,\pi^-$) showing the $\Sigma^+$ signal
  together with the misidentification background \cite{Sieb10}. The vertical dashed lines represent the $3\sigma$
  cuts applied to select the $\Sigma^+\pi^-$ events. }
  \label{Sigma+}
\end{figure}        
All the charged particles in the final state are identified, and the missing mass to
K$^+$, p and $\pi^{\pm}$ allows to tag the $\Sigma^{\mp}$ as it is visible in Fig.~\ref{Sigma+} 
 \cite{Sieb10}. The missing mass to all charged particles shows a clear signal 
corresponding to the neutron that can hence be
selected \cite{Sieb10}. This way, a semi-exclusive analysis could be carried out. 
\begin{figure}
  \includegraphics[height=.60\textheight]{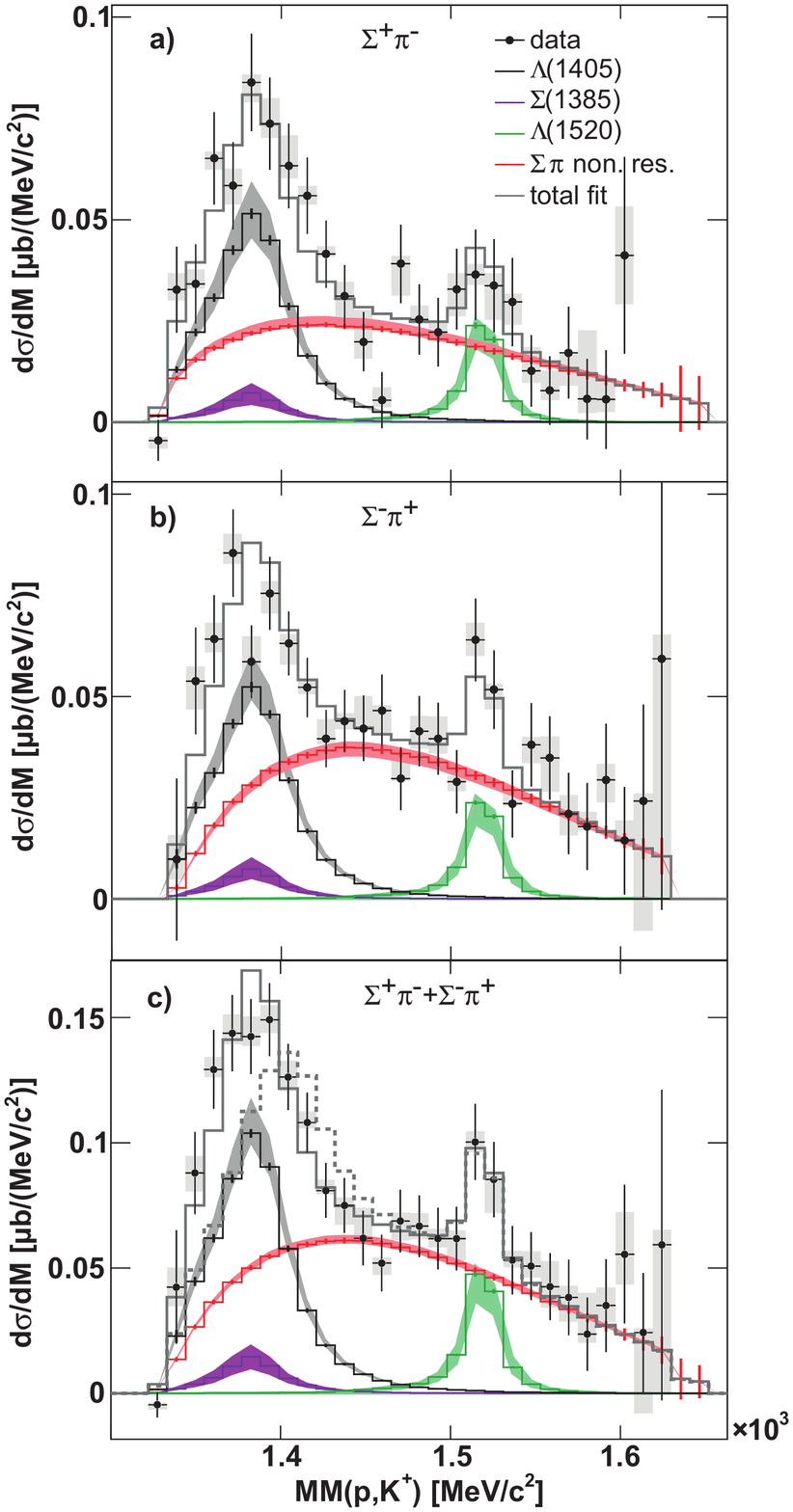}
  \caption{(Color online) Missing mass $MM(p,K^+)$ distributions for
events attributed to the  $\Sigma^+\pi^-$ decay channel a) and to the $\Sigma^-\pi^+$ decay
channel b). Panel c): sum of both spectra from panels a) and b) \cite{L1}.
The gray dashed histogram shows the sum of all simulated channels
if the $\Lambda(1405)$ is simulated with its nominal mass of $1405$ MeV/c$^2$.
Coloured histograms in the three panels indicate the contributions of
the channels (1-5) obtained from simulations. Data and simulations
are acceptance and efficiency corrected. The grey boxes indicate systematic
errors.}
  \label{L1405}
\end{figure}
\begin{figure}
  \includegraphics[height=.25\textheight]{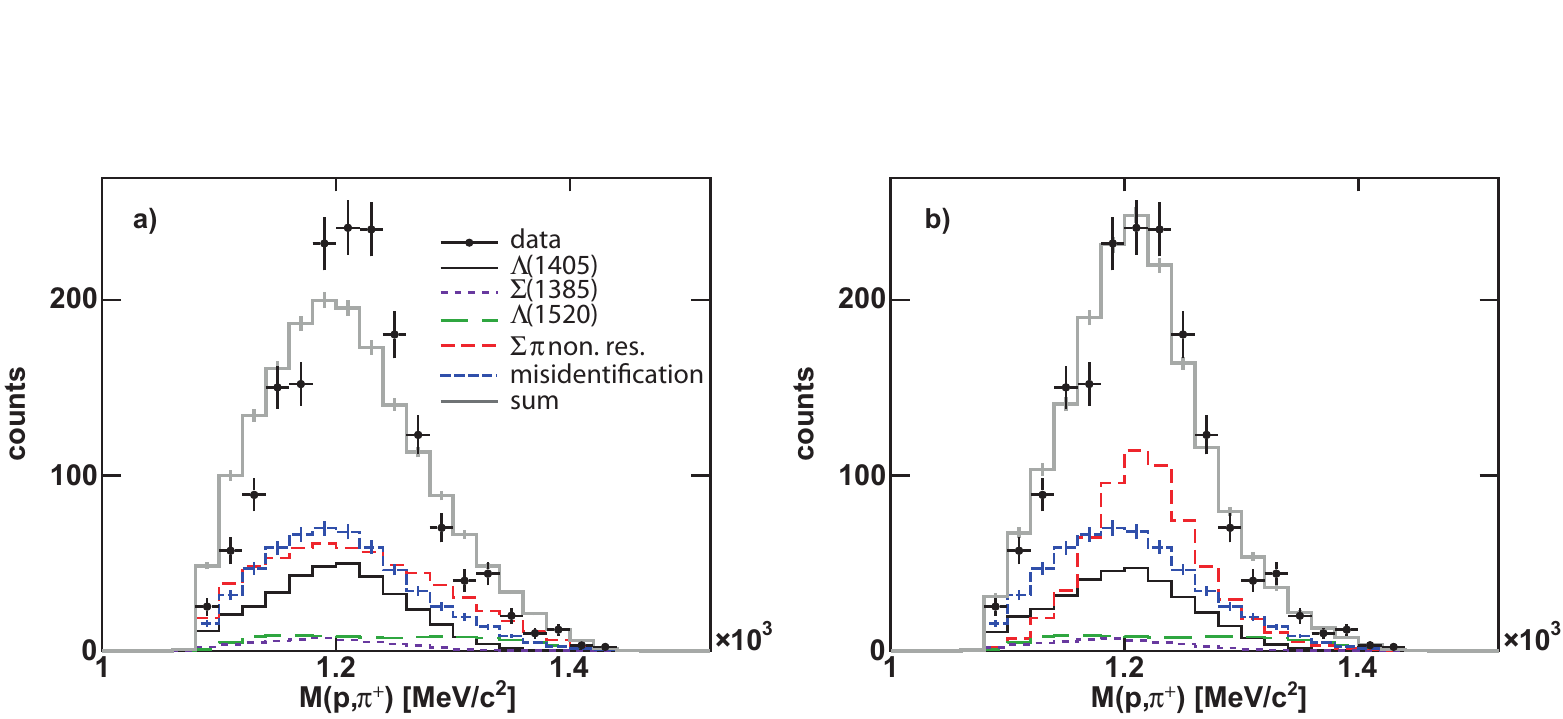}
  \caption{(Color online) p$\pi^+$ invariant mass distribution obtained after the $\Sigma^-$  
  and $K^+$ selection within the HADES acceptance \cite{D_L1405}. 
  Panel a): a phase space distribution is assumed for the $\Sigma^- \pi^+$ non-resonant part. Panel b): the
  reaction $p+p \rightarrow \Sigma^- +K^+ +\Delta(1232)^{++}$ is used to simulate 
  the $\Sigma^- \pi^+$ non-resonant part.}
  \label{L1405Back}
\end{figure}
The $\Lambda(1405)$ spectral shape is studied by means of the missing mass to the K$^+$-p pairs.
Figure~\ref{L1405} shows the obtained distributions where one can see the contribution by the non 
resonant channels $p+p \rightarrow \Sigma^{\pm} + \pi^{\mp} +K^+ +p$ .

 The data interpretation is carried out by assuming that interferences are occurring among
the intermediate states and the observed shift of the $\Lambda(1405)$ spectral shape towards lower masses
with respect to the nominal value has been discussed in \cite{L1,L2}.
The aspect of this analysis that should be underlined here is the study of the channel 
$p+p\rightarrow \Sigma^{-} +\pi^+ +p +K^+$, non-resonant in the $\Sigma^-\pi^+$ state.
This contribution is clearly visible in panel b of Fig.~\ref{L1405} (red histogram).
By looking at the p$\pi^+$ invariant mass distribution for all the selected events, the experimental data
are not reproduced correctly by simulation assuming a phase-space distribution of the channel
$p+p\rightarrow \Sigma^{-} +\pi^+ +p +K^+$. The agreement improves consistently if the reaction 
$p+p\rightarrow \Sigma^{-} +\Delta(1232)^{++} +K^+$ is considered for the total yield of the non resonant
background in the $\Lambda(1405)$. The comparison between the two assumptions is shown in the
left and right panel of Fig.~\ref{L1405Back} \cite{D_L1405}, where the p$\pi^{+}$ invariant mass
 experimental distribution  is compared to simulations including the phase-space background 
 distribution (left panel) and a simulation that contains the $\Delta(1232)^{++}$ resonance (right panel).
 It is clear that the experimental data favour the $\Delta(1232)^{++}$ hypothesis.
 
For the charge conjugated channel $p+p\rightarrow \Sigma^{+} +\pi^- +p +K^+$ no resonance is clearly visible
in the p$\pi^-$ invariant mass (see Fig. 2 in\cite{D_L1405}). In this case the formation 
of one or more rather broad N$^*$ resonances 
could occur but it would not be easy to detect them in the invariant mass spectrum because of their large widths.
On the other hand one can not exclude that also in this case the resonance formation replaces the
phase-space emission of the same final state.\\

\begin{figure}
  \includegraphics[height=.25\textheight]{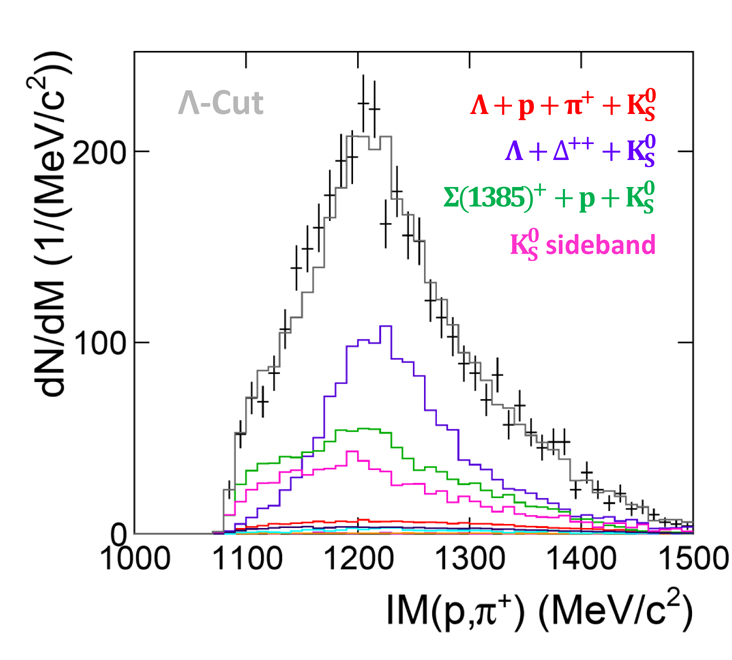}
  \caption{(Color online) p$\pi^+$-invariant mass distribution after the $\Lambda$-cut on the missing mass 
  distribution to the identified p,$\pi^+,\,\pi^+,\,\pi^-$ and with a cut on the $K^0_S$ mass in the
   $\pi^+\pi^-$-invariant mass spectrum \cite{K0spp}. The grey histogram corresponds to the sum of simulated contributions
    together with the background defined by the sideband sample.}
    \label{K0sDeltaLambda}
\end{figure}

\begin{figure}
  \includegraphics[height=.25\textheight]{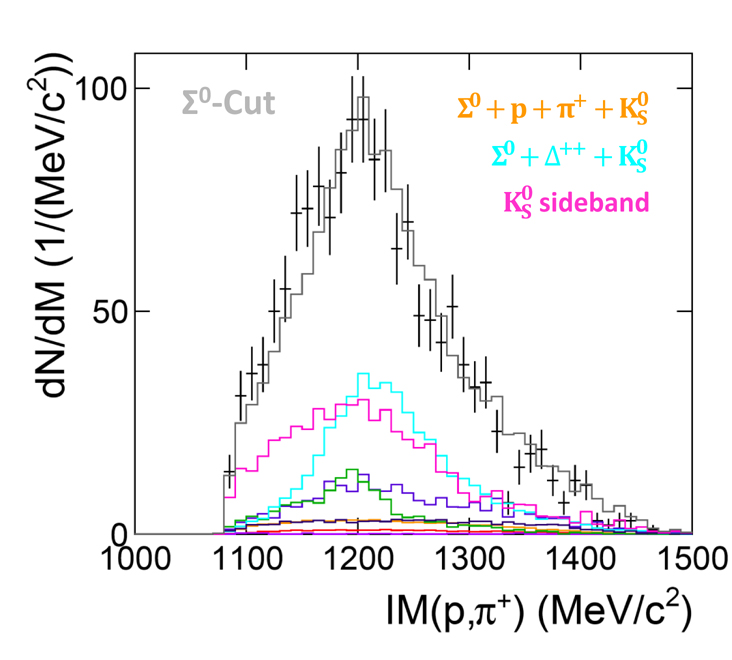}
  \caption{(Color online) p$\pi^+$-invariant mass distribution after the $\Sigma$-cut on the missing mass 
  distribution to the identified p,$\pi^+,\,\pi^+,\,\pi^-$ and with a cut on the $K^0_S$ mass in the
   $\pi^+\pi^-$-invariant mass spectrum \cite{K0spp}. The grey histogram corresponds to the sum of simulated contributions
    together with the background defined by the sideband sample.}
    \label{K0sDeltaSigma}
\end{figure}

A similar case is discussed in the analysis of $K^0_S$ produced in p+p collisions at $3.5$
 GeV \cite{K0spp}, where the formation of a $\Delta^{++}(1232)$ resonance accompanying the $K^0_S$ production
 is strongly dominant with respect to the phase-space emission of the p$\pi^{+}$ pairs if the 
 following reactions are compared:\\
 $p+p\rightarrow \Delta^{++} + K^0_S + \Lambda/\Sigma$,\\
 $p+p\rightarrow p+\pi^+ + K^0_S + \Lambda/\Sigma$.\\
Again the p$\pi^+$ invariant mass reconstructed for events  that contain additionally a $K^0_S$ and a hyperon
 allows to draw quantitative conclusions and extract production cross sections.
 Without going into the details of the exclusive analysis carried out to extract the cross section of the different
 production channels \cite{K0spp}, the extracted p$\pi^+$ invariant mass for the reactions containing a $K^0_S$
 together with a $\Lambda$ or $\Sigma$ hyperon are shown in Figs.~\ref{K0sDeltaLambda} and 
 \ref{K0sDeltaSigma} respectively. 
 The experimental distributions obtained after the kaon and hyperon selections are compared to simulation
 including the contribution of the $\Sigma(1385)$ and the associate production of the $\Delta(1232)^{++}$
 summed to the experimentally determined sideband background. The blue dashed histogram in 
 Fig.~\ref{K0sDeltaLambda} and the green dashed histogram in Fig.~\ref{K0sDeltaSigma} show the peak
 corresponding to the nominal mass of the $\Delta(1232)^++$. The dashed red lines present in both Figs.
 represent the phase space simulation of the reaction $p+p\rightarrow \Lambda/\Sigma +p+\pi^+ +K^0_S$
 and the yield of these distributions is determined by a multi-variables fit to the experimental data \cite{K0spp}.
 It is clear from these distributions that the experimental data favour the creation of an intermediate $\Delta^{++}$
 resonance with respect to the phase space emission of the p$\pi^+$ pairs; quantitatively the cross section is found
 to be 10 times higher.
These first two examples indicate that whenever possible the resonance production dominates with respect
to the phase space emission of hadrons in reactions where strange hadrons were studied.

\section{Interferences among resonances}
So far the shown analyses were carried out neglecting possible interferences among the contributing channels.
As for the $\Lambda(1405)$ case, the interference scenario has been studied in \cite{L2} but
it leads to unrealistic cross sections of the different channels. As for the $K^0_S$ analysis reported in \cite{K0spp} 
and discussed above, most of the investigated final states contain four or five hadrons, such that the effect
of interference might not be strongly visible experimentally. 
On the the other hand, it has been observed by studying the reaction: $p+p \rightarrow p+K^+ +\Lambda $ at $3.5$ GeV
that not only a large fraction of the total yield is connected with the intermediate production of several N$^*$ states 
($p+p\rightarrow p+N^+(1650,\,1710,\,1720,\,1800,\,1850,\,1900\,\mathrm{MeV/c^2}$) as already discussed 
in \cite{Cos10}, but also a clear signature of interference effects has been measured. 
The study of the interferences has been possible employing a Partial Wave Analysis 
(PWA) of the final states utilising the Bonn-Gatchina framework \cite{BG01,BG02}. 
\begin{figure}
  \includegraphics[height=.6\textheight]{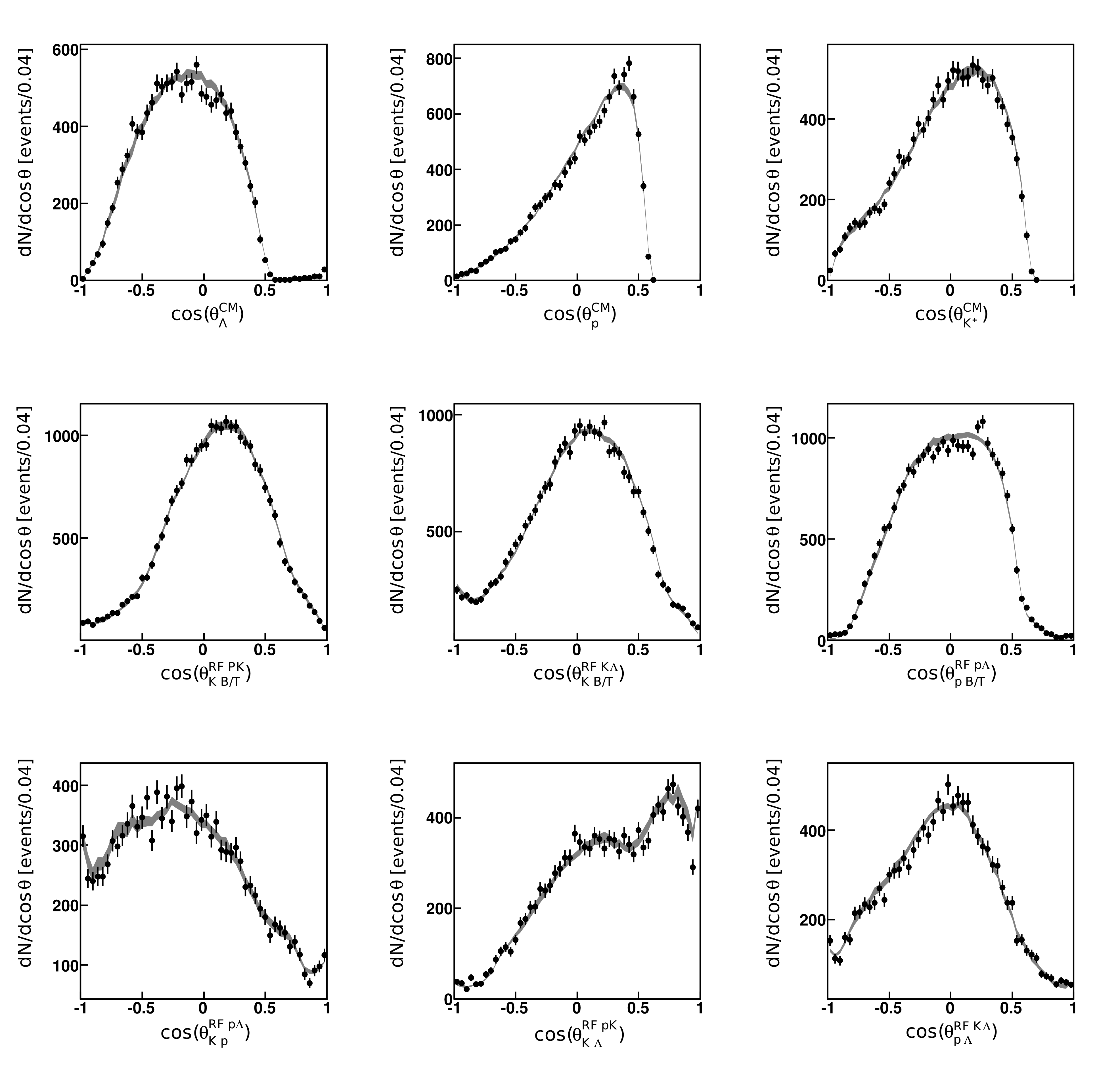}
  \caption{Angular correlations for the $pK^+\Lambda$ final state, within the detector acceptance
  , shown for the HADES data-set. Black dots are the experimental data with their statistical 
  uncertainty while the grey band shows the four best solutions of the PWA and displays their
   systematic differences. The upper index at the angle indicates the rest frame (RF) in which
    the angle is displayed. The lower index names the two particles between which the angle is evaluated. 
    CM stands for the center-of-mass system. B and T denote the beam and target vectors, respectively.
     The observables are: CMS angles (upper row), Gottfied-Jackson angles (middle), and helicity angles (lower row).}
    \label{PWAAngle}
\end{figure}

Figure \ref{PWAAngle} shows the final result of the PWA analysis of the HADES data 
for the reaction $p+p\rightarrow p+K^+ +\Lambda $ at $3.5$ GeV \cite{PWAHad}. The data
points within the HADES acceptance are shown together with the result of an event-by-even
partial wave fit, where several N$^*$ resonances ($N(1650)\frac{1}{2}^-, \,N(1710)\frac{1}{2}^+, 
\,N(1720)\frac{3}{2}^+, \,N(1875)\frac{3}{2}^-,$ $\,N(1880)\frac{1}{2}^+$$,\, 
N(1895)\frac{1}{2}^-$and $N(1900)\frac{3}{2}^+$) and non-resonant configurations 
with different quantum numbers were used as hypotheses.  
One can see that the PWA manages to model the experimental data much better than
the attempt shown in \cite{Fab13} where the incoherent sum of similar resonances has been employed.
As discussed in \cite{PWAHad} the grey band shown in Fig.~\ref{PWAAngle} represents
the four best solutions extracted from the PWA, which means that it has been not possible to determine
the precise content of the experimental data by applying this kind of analysis to only one 
data set.
\begin{table}[h]
\begin{tabular}{llcccccccc}
experiment &	energy (GeV)	& momentum (GeV/c)	& $\sqrt(s)$ (GeV)	& excess energy (pK$\Lambda$) (MeV)& 	max. N*  mass (MeV)& 	excess energy (ppK$^-$) (MeV) &	statistic	& polarization \\
\hline
COSY-TOF	& 1,82 &	2,59	&2,63 &	84,87 &	1694,23 &	-231,40 & 	791 &	N \\
\hline
COSY-TOF &	1,90&	2,68	&2,66&	114,91	&1724,27&	-201,35&	1037&	N\\
\hline
COSY-TOF&	1,92&	2,70	&2,67&	121,56&	1730,92	&-194,71	&	160000	&?\\
\hline
COSY-TOF &	2,06&	2,85 &	2,72	 & 171,05	& 1780,41 & 	-145,22 &	4323	&N\\
\hline
DISTO &	2,15 &	2,94	& 2,75 &  	200,44 	& 1809,80 &	-115,83	 &	121000	 & Y\\
\hline
COSY-TOF	& 2,16	& 2,95	& 2,75	& 203,69 &  	1813,05  &	-112,58	 &	43662	& Y\\
\hline
COSY-TOF	& 2,16  &	2,95  &	2,75 &	203,69	  &  1813,05 &	-112,58	   & 	7228	& N\\
\hline
COSY-TOF	& 2,16 & 	2,95&	2,75&	203,69	 &1813,05&	-112,58	  &	15372	&N\\
\hline
COSY-TOF	&2,26	&3,06&	2,79	&238,95	&1848,31&	-77,32&	89684	&N\\
\hline
COSY-TOF	&2,28&	3,08&	2,79&	245,70&	1855,06&	-70,57&	30000&	N\\
\hline
COSY-TOF&	2,40	&3,20&	2,83&	284,06	&1893,42&	-32,21&	3322&	N\\
\hline
COSY-TOF	&2,40&	3,20&	2,83&	284,06&	1893,42&	-32,21&	5791&	N\\
\hline
COSY-TOF&	2,50&	3,31&	2,87&	318,86&	1928,22&	2,60&	6263	&N\\
\hline
DISTO	&2,50	&3,31&	2,87&	318,86&	1928,22&	2,60&		304000&	Y\\
\hline
DISTO	&2,85	&3,67	&2,98	&430,48	&2039,84&	114,21	&	424000&	Y\\
\hline
FOPI	&3,10	&3,93&	3,06&	508,97&	2118,33&	192,70&		1000&	N\\
\hline
HADES&	3,50&	4,34	&3,18&	629,33&	2238,69&	313,06	&	20000&	N\\
\hline
\end{tabular}
\caption{Summary table of the available data $p+p\rightarrow p+K^+ + \Lambda$ reaction.}
\label{PWATab}
\end{table}
Also, this analysis showed us that polarisation observables would help in the precise determination
of the different N$^*$ waves. For this reason, we plan to apply the PWA analysis to
all the data set measured at different kinetic energies in the GeV energy range 
and available to date. This should 
better constraint the results and obtain an excitation function for the different N$^*$.
Several measurements of the reaction $p+p \rightarrow p+K^++\Lambda$ in a fixed target configuration 
have been carried out by the DISTO and COSY-TOF collaborations \cite{CT} in the past 
and some of these data contain also polarisation
observables. The available data are summarised in Table \ref{PWATab} together with information
about the kinetic energy, statistics and availability of spin observables.
 One can see that the proton kinetic  energy varies
between $1.8$ and $3.5$ GeV. A global PWA of these data can constrain the excitation functions of the different N$^*$
with a mass between $1.7$ and $2$ GeV/c$^2$ with unprecedented precision and these data can be used as a
reference of the upcoming experiments at FAIR at higher kinetic energies. 

The PWA considered here interests the strange final state of N$^*$ resonances, but analogous studies can be carried out
in the one or two pions final states ($N^*\rightarrow N\pi/N\pi\pi$). 
In \cite{Ad14} it is shown for the first time how for the same p+p at $3.5$ GeV experiment the relative
 contribution of N$^*$  and $\Delta$ resonances to the non-strange final states
and then to the di-electron invariant mass spectrum has been determined within an extensive resonance model. This model
includes appropriate angular distributions for the different contributing channels, partially extracted from exclusive measurements
 but it assumes an incoherent sum for all the contributing channels. In order to understand the role played by interferences 
 in non strange final states and reconcile all the HADES analysis, a PWA has been already carried out for the non-strange
 final states at lower kinetic energies first \cite{Wit13}. There the effect of resonances manifests itself clearly
 but a consistent picture must still be worked out.
 This kind of analysis will be soon extended also to the $3.5$ GeV data.
 
 It is clear that whatever effect the interference among resonances might have on the final spectra
  in p+p and n+p collisions, current
 transport calculations for such colliding systems do not include this mechanism at all. As a matter of fact
  for some probes, as for example the $\Lambda$ hyperons, yield and angular distribution
  are not correctly modelled in the transport calculations in p+p and 
 p+Nb \cite{LamPNb} collisions at $3.5$ GeV. 
 Going a step further, one can ask whether the coherence of the resonance emission is completely broken
 when moving from p+p/p+n to p+A and A+A collisions or if the effect of the interference is persistent also 
 there, at least to some extent. For the specific case of the $\Lambda$ hyperon $p_T$ and rapidity 
 distributions might be washed out by scattering of the nucleons in p+A and A+A collisions,
  but since the interferences also influence cross sections, the effect could still be sizeable.    
  Provided that A+A collisions can be seen as the superposition of multiple nucleon-nucleon collisions, interferences
  could play a role in the total yield of some hadrons. But assuming that hadronisation follows the
  creation of an intermediate state of matter \cite{Gal12} in A+A collisions already in the GeV energy range 
  the picture could be completely different and the effect of interferences completely negligible.
  It is clear that a better understanding of these processes is needed.
  
\section{The case of heavy non-strange resonances}
The last but not least point of the resonance discussion in the GeV energy range
deals with the production of heavy non-strange resonances coupling to strange final states.
This story begins with the observation of the $\Sigma(1385)^+$ in p+p collision at $3.5$ GeV 
where the reaction  $p+p\rightarrow \Sigma(1385)^+ (\rightarrow \Lambda+ \pi^+)+n+K^+ $ \cite{S1385+} 
has been investigated.
This reaction has been measured semi-exclusively by detecting all the charged particles in the
final state, reconstructing the intermediate $\Lambda$ hyperon via its decay into p-$\pi^-$ pairs
and by applying the missing mass selection for the undetected neutron.
\begin{figure}
  \includegraphics[height=.25\textheight]{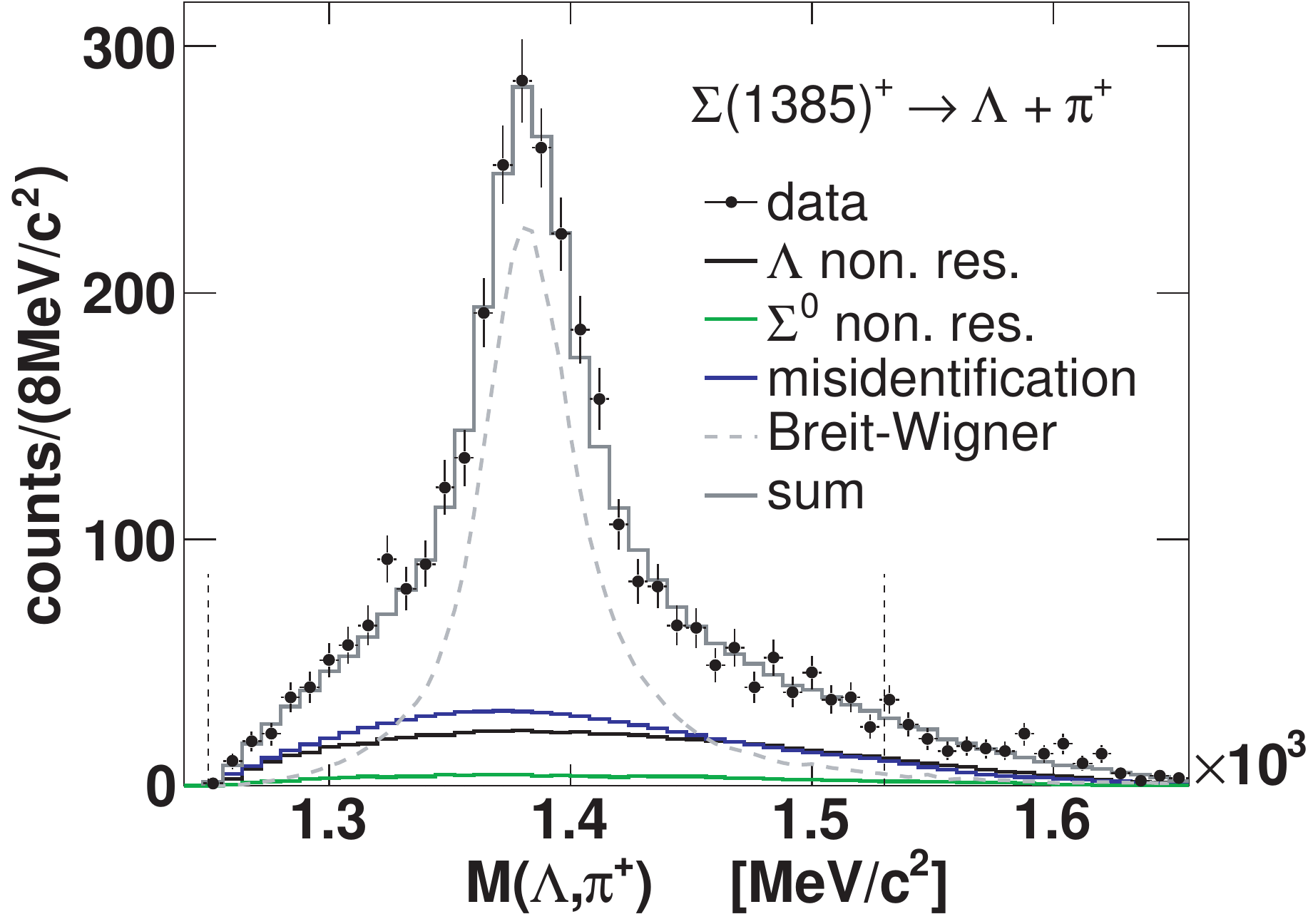}
  \caption{(Color online) Invariant mass distribution of the $\Lambda-\pi^+$ pairs \cite{S1385+}.
  The circular symbols show the experimental data, the black and green curve depict the contribution
  by the non-resonant production, the blue curve shows the estimated misidentification
  background and the grey dashed curve represents a Breit-Wigner function fitted to the data. 
  The solid grey line displays the sum of the the simulated contributions fitted to the data.}
  \label{S1385}
\end{figure}
\begin{figure}
  \includegraphics[height=.3\textheight]{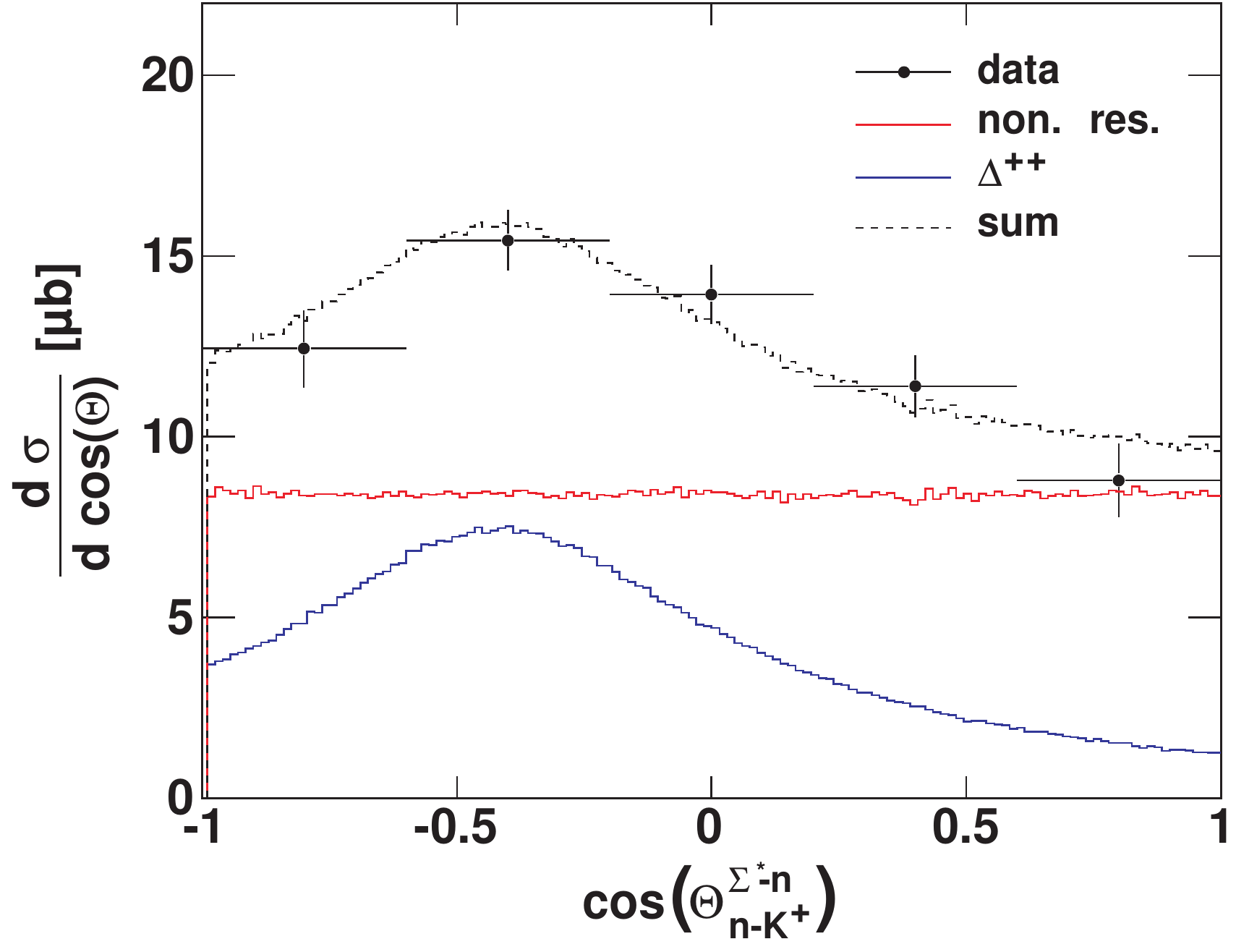}
  \caption{(Color online) Differential cross section of the $\Sigma(1385)^+$ as a function of 
  $\cos \left ( \Theta ^{\Sigma^*-n}_{n-K^+}\right)$ \cite{S1385+}. The red and blue histograms show the
  simulated contribution for a non-resonant production of the $\Sigma(1385)^+$ and via an
  intermediate $\Delta^{++}$, respectively.}
  \label{SigmaAngle}
\end{figure}
 Figure \ref{S1385} shows the $\Lambda\pi^+$ invariant mass distribution where the $\Sigma(1385)^+$
peak is recognisable over a moderate unphysical background originating from the wrong identification
of K$^+$ and a physical, non resonant background ($p+p\rightarrow \Lambda + \pi^++n+K^+ $).
The signal strength has been extracted differentially with respect to the particle emission angle
by fitting the invariant mass distributions with a modified Breit-Wigner function in different 
bins of the angle variable. Different choices of the angular variables are possible
as discussed in \cite{S1385+} where the differential cross sections are calculated
as a function of the p-p center of mass angle, helicity and Gottfried-Jackson angles.

In order to extract
a differential cross section for the $\Sigma(1385)^+$ production acceptance and efficiency 
correction must be applied. This correction procedure requires simulations that include 
a model for the $\Sigma(1385)$ production, in order to extrapolate correctly the yield in the phase 
space region that is not covered by the acceptance of the HADES spectrometer.
An iterative method is employed, as described in \cite{Sieb10}, assuming a certain production mechanism
in $4\pi$ for the $\Sigma(1385)$ and correcting the data with the simulation hypothesis.
The angular distribution of the corrected data are then compared to the model
and in case of deviations the model is modified and the procedure repeated.

This way, it was found that at least $30\%$ of the $\Sigma(1385)^+$ yield stems from the decay
of a resonance with a mass of around $2000$ MeV/c$^2$ indicated by us as a $\Delta(2000)^++$
following the observation reported in \cite{Chi68}. The HADES data do not allow for a more exclusive
analysis of the hypothetical heavy resonance nor the determination of its quantum numbers.
Still, even though the available phase space for the production of a resonance with a mass larger than
$2$ GeV/c$^2$ is limited in p+p reactions at $3.5$ GeV, the contribution to the $\Sigma(1385)^+$
seems relevant.

  Among the most relevant results by the HADES collaboration, non-trivial in-medium effects for both
  strange and non strange final states have been observed in Ar+KCl collisions at $1.756$ AGeV.
  These results divide themselves in measured excesses of particle yields with respect to either experimental
  references \cite{Hol11} or calculation \cite{XiAr} and observed shifts of momentum distribution with respect to transport
  calculations \cite{K0sArKcl,K0pp}. 
 The deep sub-threshold production of the doubly strange hyperon $\Xi^-$ in Ar+KCl reactions in a fixed target configuration
 at kinetic energies of $1.76$ AGeV showed that the $\Xi^-/(\Lambda+\Sigma^0)$ ratio is significantly larger than the number
 obtained from available predictions \cite{XiAr} based on statistical hadronization models. 
Taking the NN threshold of the $\Xi$ production at $\sqrt{s}=\,2.5$ GeV it is clear that secondary collisions and eventually in-medium
 modification of the vacuum hadron properties could play a role here.
 The  $\Xi^-/(\Lambda+\Sigma^0)$ ratio is found to be equal to $(5.6 \pm 1.2 ^{+1.8}_{-1.7})\cdot 10^{-3}$ 
 and more than a factor $20$ larger than the
 value extracted from a statistical hadronization model applied to Au+Au collisions at $1.76$ AGeV.
 One has to observe that all other hadron species measured in the Ar+KCl experiment agree with the
 value predicted by the thermal model. 
 This excess could so far not be explained, even including the contribution of reactions like $Y+Y \rightarrow \Xi + N$
  in adequate transport calculation that model the heavy ion reactions \cite{Kol12,Theo1Xi}. This measurement speaks in favour
  of the formation of an intermediate form of matter beyond a trivial superposition of nucleon-nucleon collisions
  followed by pion-nucleon and $\Delta$-nucleon secondary reactions. Indeed, in case of a deconfined phase
  of matter the production of double strange baryon could proceed easier in a sort of catalytic 
  mechanism as discussed in \cite{Gal12}.
  
  On the other hand if one looks at recent results from p+A reaction at $3.5$ GeV \cite{XipN}, a similar unexpected
  excess for the $\Xi^-$  production probability is found if one compares the measured value of
  $2.0 \pm 0.4 (\mathrm{stat}) \pm 0.3 (\mathrm{norm}) \pm 0.6 (\mathrm{syst})) × 10^{-4}$
  to model prediction assuming a statistical hadronization.
  Provided that there is not real reason to believe that the hadronization happens governed by
  any thermal equilibrium in p+A collisions in the GeV energy region, we still deal here with a sub threshold
  production for the $\Xi^-$.
  
  The question can be asked whether also in this case as for the $\Sigma(1385)^+$ production in p+p
  reactions at $3.5$ GeV the $\Xi^-$ yield might stem from the decay of a massive strange or non-strange
  resonances in processes as $\Lambda(>2000) \rightarrow \Xi^-+K^+$ or $\Delta^+(> \,2350 )\rightarrow \Xi^- + K^+
  +K^+$. This possibility cannot be tested directly with the available data in p+p and p+Nb collisions exploring
  the same method used for the analysis of the $\Sigma(1385)^+$ because of the reduced statistics but also
  because p+A collisions do not allow for exclusive analysis. This hypothesis could be investigated in high
  statistics p+p collisions with kinetic energies above the N-N threshold of $3.7$ GeV as it will soon possible 
  with HADES. The improved data acquisition system and particle identification capability for
   positive and negative kaons via the RPC detector can help to address this kind of decay via exclusive analyses in the near future. 
 
 Massive non-strange resonances could obviously also be produced in A+A collisions already at energies
 below the NN threshold by multiple collisions and considering the higher tails of the Fermi distribution
 of the momentum distribution of the single nucleon. Which means that the excess measured in Ar+KCl
 for the $\Xi^-$ baryon could also stem from the decay of an heavy resonance and not formed
 in a pre-hadronic state of matter. 
This scenario can be tested in elementary hadron-hadron collisions and then included in the modelling
of heavy ion collisions to be checked against other hypotheses.

Also for the non-strange sector similar mechanisms coupled to resonances could occur.
In particular, one can test hypotheses where heavy N$^*$ resonance decay into a final state
containing a $\eta$ or $\rho / \omega$ meson. An excess has been measured in the $e^+-e^-$ invariant mass 
spectra collected for the Ar+KCl reaction at $1.76$ AGeV with respect to an adequately normalised
 reference spectrum measured in p+p and p+n collisions at slightly lower  ($1.25$ AGeV) kinetic energies \cite{Hol11}.
 
 The excess in the dilepton yield is measured in the  $e^+-e^-$ invariant mass region right below the 
 $\rho$ nominal mass in correspondence of the region where the decay of $\eta$ in the dilepton final
 state plays and important role.
 These results have been interpreted by theory either advocating in-medium modification of the $\rho$ spectral
 shape \cite{Wei14} or the contribution of the decay of $\Delta$ resonances \cite{Bra13}. It is natural to ask to which 
 extent the coupling of resonances to final states containing dileptons influences the invariant mass
 region where the excess is observed and also whether objects heavier than $1.5$ GeV/c$^2$ play a role here.
 
 Exclusive measurements of the reaction $p+p-> \eta/\omega +p+p$  have been carried out 
 by measuring the two protons in the final state, employing the missing mass technique to 
 select the $\eta$ and $\omega$ mesons and finally applying a kinematic refit to improve the
 momentum resolution \cite{Kahl}. The p$\eta/\omega$ invariant mass distributions
 obtained after the selection of the two final states can be used to look for the signature left by the 
 presence of resonances and in the case of the $\eta$ meson about $47\%$ of the total yield 
 has been attributed to the the N$^*(1535)$ \cite{Kahl}. For the $\omega$ channel the data seem 
 in agreement with a phase space production.
  This kind of measurements allows also to investigate whether
 the angular distributions of the $\eta+p+p$ and $\omega+p+p$ shows any signs of a resonance
 decaying into a meson-proton pair ($N^{*+}\rightarrow \eta/\omega+p$). 
 As discussed in the $\Sigma(1385)^+$ analysis \cite{S1385+} different angles can be defined for  
 p+p collisions with a three body final state. 
As it is shown in Fig.~\ref{SigmaAngle} the helicity distribution with respect to $\Theta^{\Sigma^*-n}_{n-K^+}$,
i. e. the angle between the $\Sigma(1385)^+$ and the neutron in the neutron-K$^+$ reference system,
shows a different pattern for the non-resonant $\Sigma(1385)^+$ emission with respect to the case
$\Delta(2000)^{++}\rightarrow \Sigma(1385)^+ +K^+$. The same variable can be constructed 
for the $\eta$ and $\omega$ missing mass analyses where then the helicity angles will be
$\Theta^{p\eta}_{pp}$  and $\Theta^{p\omega}_{pp}$, angles between the $\eta/\omega$ and the proton in the final
state in the reference system of the two final state protons. It is clear that since the two protons in the final
state are indistinguishable, both permutations have to be considered. 
 \begin{figure}
  \includegraphics[height=.3\textheight]{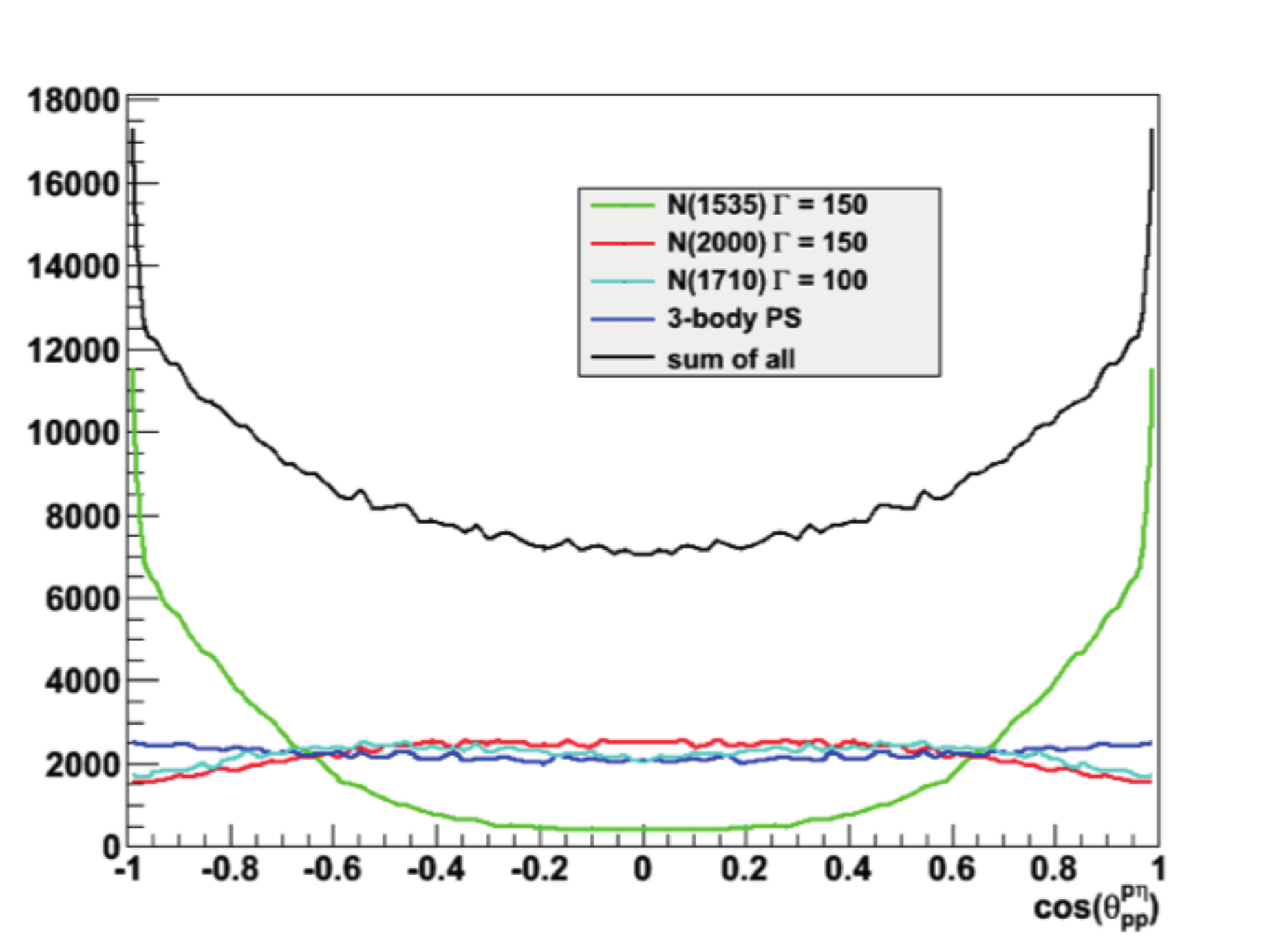}
  \caption{(Color online) Simulated distribution of the $\cos \left( \Theta^{p\eta}_{pp}\right)$ for the final state 
  $p+p\rightarrow p+p+ \eta$. The dark blue line represents a phase space emission, the green, red and cyan lines
  correspond to simulation including the intermediate production of a N$^{*+}$ with a mass of $1535, \,1710$ and $2000$ MeV/c$^2$ 
  respectively. The black line represents the sum of all the simulated contributions.}
  \label{etaHelAngle}
\end{figure}
Figure~\ref{etaHelAngle} shows the results from simulations of the reaction $p+p\rightarrow p+p+ \eta$ assuming
either a phase space emission of the three particles in the final state or the production of an intermediate
N$^{*+}$ resonance with a mass of $1535, \,1710$ and $2000$ MeV/c$^2$. These simulations represent the distribution
in the full phase space and do not account for the detector acceptance or reconstruction efficiency and each contributions
has the same weight in the total sum.
It is clear from the distributions that the helicity angle measured in case of a massive resonance changes significantly with 
respect to the angular distribution characterising a phase space emission. In particular, a difference is also visible if one 
compares heavier and lighter N$^{*+}$ resonances. A similar distribution as shown in Fig.~\ref{etaHelAngle} is 
obtained when simulating the reaction  $p+p\rightarrow p+p+ \omega$.
The effeciency of the selection of the resonance contribution via the helicity angle variable should be
tested after having considered the effect of the geometrical acceptance and reconstruction efficiency of the
measured final state in the HADES spectrometer. 
\begin{figure}
  \includegraphics[height=.3\textheight]{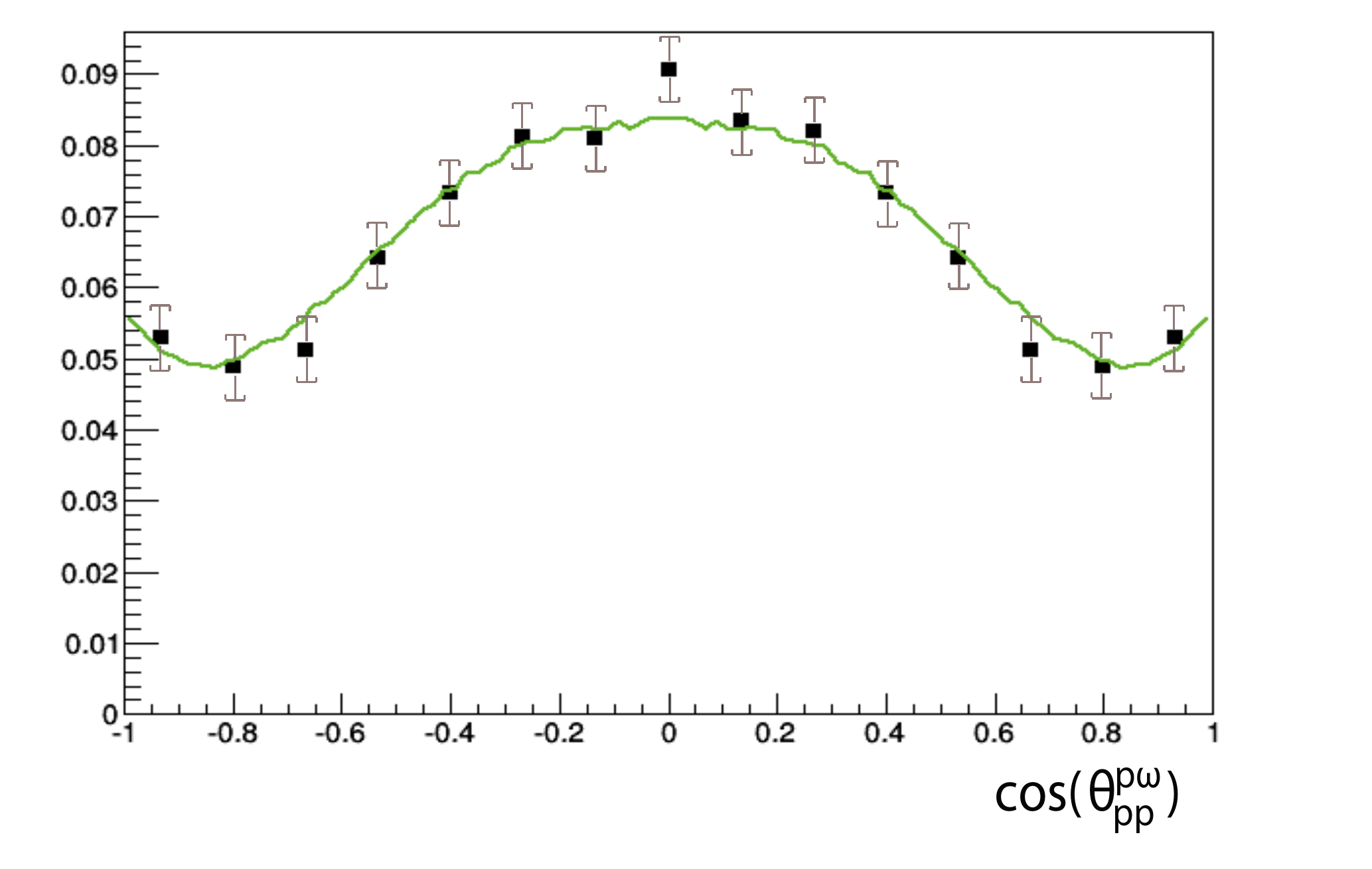}
  \caption{(Color online) Experimental distribution of the $\cos \left( \Theta^{p\omega}_{pp}\right)$ for the final state 
  $p+p\rightarrow p+p+ \omega $ (squared symbols) together with the distribution of the same helicity angle obtained 
  from phase-space full scale simulations. Errors represent an estimation of the systematics.}
  \label{omHelAngle}
\end{figure}
Figure \ref{omHelAngle} shows the $\cos \left( \Theta^{p\omega}_{pp}\right)$ distribution for the reaction
$p+p\rightarrow p+p+ \omega$. The black symbols represent the experimental data reconstructed within the
HADES acceptance and the green histogram shows the results of full scale simulations, which hence take
into account the precise response of the HADES spectrometer, for an event generator assuming the phase space
emission of the considered reaction.
It is clear that the geometrical acceptance selects a region of the phase space where the different resonance-scenarios
display very similar distributions in the helicity angle. In particular, one can see that the experimental data are well reproduced
by the phase-space simulations for the $\omega$ channel as observed for the $\omega$-p invariant mass distribution
discussed in \cite{Kahl}.

New opportunities are offered by measuring similar final states with a pion in the initial state substituting the
proton beam. Then reactions as $\pi^- +p \rightarrow \eta/\omega + n$ could be addressed for the tagging of 
different resonances. In this case, the direct detection of the $\eta$ and $\omega$ mesons becomes mandatory
but  the upcoming electromagnetic calorimeter within HADES \cite{HADCal} should help the identification of channels 
as $\eta \rightarrow \gamma +\gamma $, $\eta \rightarrow \pi^0 +\pi^+ +\pi^- $ and $\omega \rightarrow \pi^0 +\pi^+ +\pi^- $.
The kinematic in the case of pion beams should be less focused towards the beam direction allowing for
a larger geometrical acceptance within HADES. After the successful commissioning of the pion beam 
at the SIS18 with the HADES spectrometer new experiments can now be planned for the near future also in
this context.
It is clear that the same hold for the investigations connected to the $\Xi$ production in elementary collisions
and the interplay with strange and non-strange resonances. Also there pion beams can be employed to study 
the exclusive final states.

\section{Summary}

The associate production of baryonic resonances together with hyperons and kaons
has been discussed and it has been shown that new exclusive analyses of the reaction
p+p at $3.5$ GeV by the HADES collaboration could pin down quantitatively
the contribution of this resonances. In the specific case of the associate of the $\Delta(1232)^{++}$
with $K^0_S$ a direct comparison with the phase space emission of the p$\pi^+$ pair
showed that the resonant mechanism is a factor 10 more probable than the non-resonant channel.
This way, the dominance of the resonances production has been verified in the studied energy regime.
The second important aspect when dealing with resonances is the effect of interferences.
One of the first PWA for p+p collisions has been carried out by the HADES collaboration to describe
the final state $pK^+\Lambda$ measured at a kinetic energy of $3.5$ GeV. This analysis shows
itself as suited to correctly reproduce the experimental data but presents ambiguities in the determination
of the specific waves contributing to the total yield. A new enterprise is presented here, aiming
to apply the same PWA published by HADES to all available data for the same reactions measured
at different energies. This project will provide a more solid base for the determination of 
the excitation function of numerous N$^*$ resonances in the mass range $1650-1900$ MeB/c$^2$.
The last aspect considered in the work is the role played by heavy resonances ( $M>\,2000$ MeV/c$^2$)
in the production of strange and non-strange hadrons not only in elementary but also in heavy ion collisions.
It has been shown that a semi-exclusive measurement of the $\Sigma(1385)^+$ allows to determine that
about $30\%$ of the total yield stems from the decay of a massive resonance here indicated as a $\Delta(2000)^{++}$.
The possibility that a similar process could be responsible for the large $\Xi^-$ yield in both p+Nb and Ar+KCl
reactions is discussed, but this scenario could not be tested quantitatively yet.
Similar considerations about the influence of massive objects to the production of $\eta$ and $\omega$
mesons are brought up, showing that p+p collisions at $3.5$ GeV do not suit the identification of such a contribution.
New perspectives offered by experiments with pion beams at GSI have been discussed as very promising.

%






\bibliographystyle{aipprocl} 




\end{document}